\documentclass{article} 
\usepackage{iclr2021_conference,times}

\usepackage{amsmath,amsfonts,bm}

\def\eqref#1{equation~\ref{#1}}

\def\1{\bm{1}}

\def\rvepsilon{{\mathbf{\epsilon}}}

\def\rvx{{\mathbf{x}}}

\def\rvz{{\mathbf{z}}}

\DeclareMathAlphabet{\mathsfit}{\encodingdefault}{\sfdefault}{m}{sl}
\SetMathAlphabet{\mathsfit}{bold}{\encodingdefault}{\sfdefault}{bx}{n}

\newcommand{\E}{\mathbb{E}}

\DeclareMathOperator*{\argmax}{arg\,max}
\DeclareMathOperator*{\argmin}{arg\,min}

\usepackage{hyperref}
\usepackage{url}

\usepackage[american]{babel}

\usepackage{mathtools} 
\usepackage{booktabs} 
\usepackage{tikz} 

\usepackage{nicefrac}       
\usepackage{amsfonts, amssymb}
\usepackage{graphicx}
\usepackage{subcaption}
\usepackage{wrapfig}
\usepackage{adjustbox}
\usepackage{enumitem}

\usepackage{microtype}
\usepackage{booktabs} 
\usepackage{adjustbox}
\usepackage{wrapfig}

\newcommand{\Enc}[2]{{q_{\phi}(#1|#2)}}
\newcommand{\HEnc}[2]{{q_{\theta, \phi}(#1|#2)}}
\newcommand{\Dec}[2]{{p_{\theta}(#1|#2)}}

\newcommand{\KL}[2]{{\mathrm{KL}\left[#1 \|#2 \right]}}

\title{Diagnosing Vulnerability of Variational Auto-Encoders to Adversarial Attacks}

\author{Anna Kuzina \\
Vrije Universiteit Amsterdam\\
\texttt{a.kuzina@vu.nl} 
\And
Max Welling \\
Universiteit van Amsterdam \\
\texttt{m.welling@uva.nl} \\
\And
Jakub M. Tomczak \\
Vrije Universiteit Amsterdam\\
\texttt{j.m.tomczak@vu.nl}
}

\iclrfinalcopy 

\begin{document}
\maketitle
\begin{abstract}
  In this work, we explore adversarial attacks on the Variational Autoencoders (VAE). We show how to modify data point to obtain a prescribed latent code (supervised attack) or just get a drastically different code (unsupervised attack). We examine the influence of model modifications ($\beta$-VAE, NVAE) on the robustness of VAEs and suggest metrics to quantify it. \footnote{The code is published at \texttt{https://github.com/AKuzina/attack\_vae}}
\end{abstract}
\section{Introduction}\label{sec:intro}

\begin{wrapfigure}{r}{0.53\textwidth}
\vskip -20pt
\centering
\includegraphics[width=0.975\linewidth]{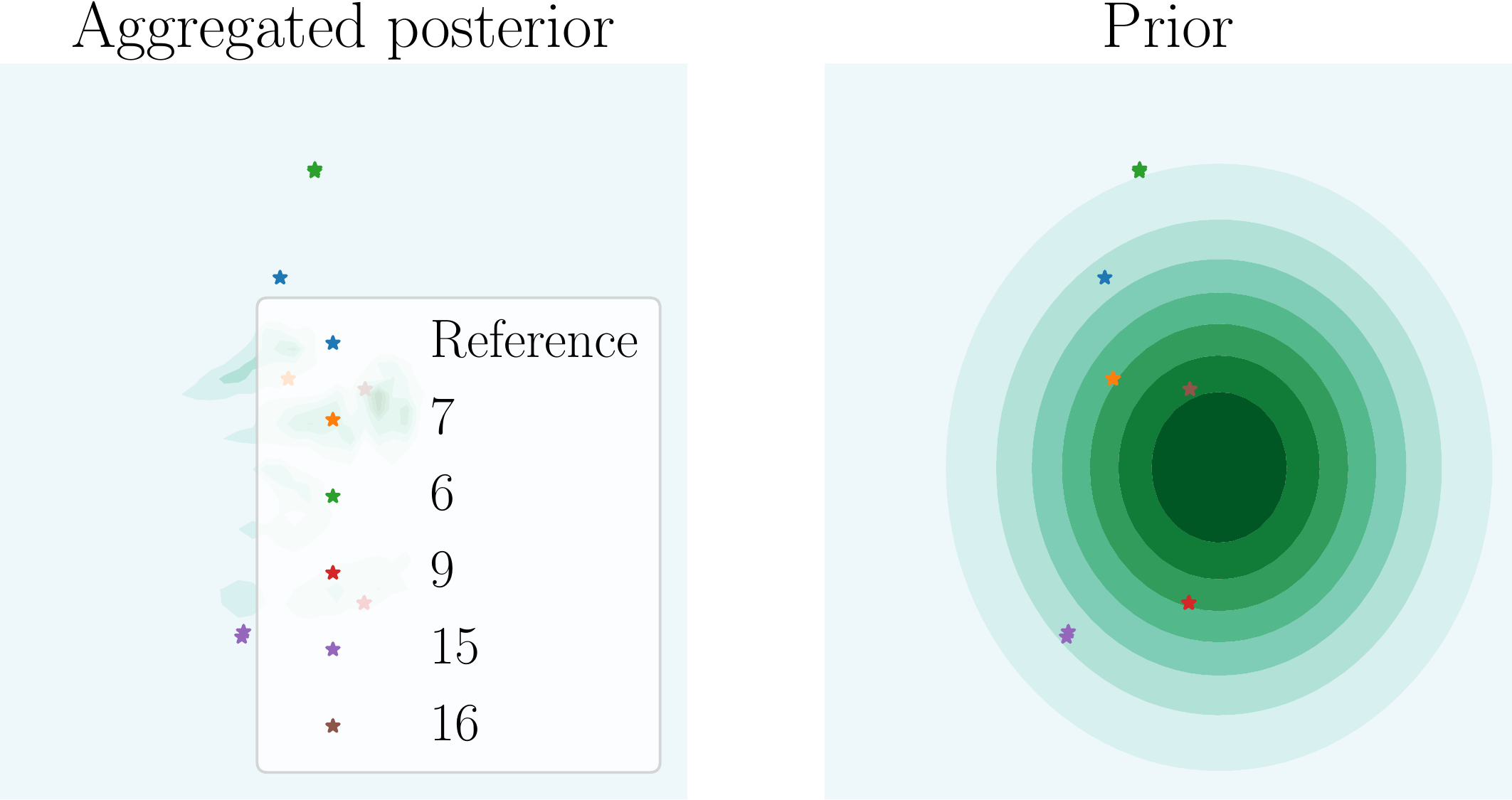}
\caption{Example of a supervised attack on VAE with 2D latent space. Given a single reference point we learn additive perturbation $\epsilon$, s.t. perturbed input has the same latent representation as a target image. We observe that a single reference point can be mapped to extremely different regions of the latent space.}
\label{fig:toy_exaple}
\vskip -15pt
\end{wrapfigure}
Variational Autoencoders (VAEs) \citep{kingma2013auto, rezende2014stochastic} are deep generative models used in various domains. Recent works show that deep Variational Autoencoders models can generate high-quality images \citep{Vahdat2020-xe, Child2020-ze}. These works employ hierarchical structure \citep{Ranganath2016-yg}, coupled with skip-connections \citep{Maaloe2019-bp, So_nderby2016-en}. An additional advantage of VAE is that it has a meaningful latent space induced by the Encoder. This motivates us to explore the \textit{robustness} of the resulting latent representations.
\textit{Adversarial attack} is one way to assess the robustness of a deep neural network. Robustness to the attacks is a crucial property for VAEs, especially in such applications as anomaly detection \citep{an2015variational, Maaloe2019-bp} or compression \citep{balle2018variational}.

\cite{Gondim-Ribeiro2018-cu} propose to minimize the KL-divergence between an adversarial and a target input to learn an adversarial attack on the standard VAE model. We show that we can use a similar strategy to attack hierarchical VAEs. \cite{Willetts2019-mu} suggest a modified VAE objective coupled with hierarchical structure as a way to increase VAE robustness to adversarial attacks. \cite{camuto2020towards} show that $\beta$-VAE tend to be more robust to adversarial attacks in terms of proposed $r$-metric. In our work, we test both $\beta$-VAE and hierarchical NVAE and show that we can attack them successfully.

In Figure \ref{fig:toy_exaple} we show an example of the \textit{supervised} attack on VAE with the 2D latent space. We show that we can add noise $\epsilon$ to a reference image so that the resulting adversarial input (second column) is encoded to a new point in the latent space. This new point is defined by the target image (first column). In \textit{unsupervised} attacks, on the other hand, we assume that the target image is not given. We show that it is still possible to construct an effective attack in this setting. The research goals of this work are the following:
\begin{itemize}[leftmargin=*]
    \item Defining robustness measures to understand how VAEs behave for adversarial attacks both in the latent space and the pixel space.
    \item Assessing robustness of hierarchical and $\beta$-VAEs to adversarial attacks.
\end{itemize}

\section{Methodology}



\subsection{Variational Autoencoders}
Let us consider a vector of observable random variables, $\rvx \in \mathcal{X}^{D}$ (e.g., $\mathcal{X} = \mathbb{R}$) sampled from  the empirical distribution $p_{e}(\mathbf{\rvx})$, and vectors of latent variables $\rvz_{l} \in \mathbb{R}^{M_{l}}$, $l=1, 2, \ldots, L$. First, we focus on a model with $L=1$ and the joint distribution $p_{\theta}(\rvx, \rvz) = \Dec{\rvx}{\rvz} p(\rvz)$. The marginal likelihood is then equal to $p(\rvx) = \int p(\rvx, \rvz) \mathrm{d} \rvz$. VAEs exploit variational inference \citep{jordan1999introduction} with a family of variational posteriors $\{\Enc{\rvz}{\rvx}\}$, also referred to as encoders, that results in a tractable objective function, i.e., Evidence Lower BOund (ELBO):
\begin{align}\label{eq:elbo}
    \mathcal{L}(\phi, \theta)
    = \E_{p_{e}(\mathbf{\rvx})}\left( \E_{\Enc{\rvz}{\rvx}}\ln \Dec{\rvx}{\rvz} - \KL{\Enc{\rvz}{\rvx}}{p(\rvz)} \right).
\end{align}
$\beta$-VAE \citep{higgins2016beta} uses a slightly modifier objective, weighting the second term by $\beta > 0$.
In case of $L>1$, we consider a hierarchical latent structure with the generative model of the following form: $p_{\theta}(\rvx, \rvz_1, \ldots, \rvz_L) = \Dec{\rvx}{\rvz_{1}} \prod_{l=1}^{L}p(\rvz_{l}|\rvz_{l+1})$, where $\rvz_{L+1} \equiv \emptyset$. There are various possible formulations of the family of variational posteriors, however, here we follow the proposition of \cite{So_nderby2016-en} where the inference model with skip-connections was proposed, namely:
\begin{equation}
    \Enc{\rvz_1, \ldots, \rvz_L}{\rvx} = \Enc{\rvz_L}{\rvx}\prod_{i=1}^{L-1}\HEnc{\rvz_i}{\rvz_{>i}, \rvx}.
\end{equation}
This formulation was used in NVAE \citep{Vahdat2020-xe}. It allows to share data-dependent information between the inference model and the generative model, because of the top-down structure.

\subsection{Adversarial attacks}
\label{sect:adversarial_attacks}
An \textit{adversarial attack} is a slightly deformed data point $\rvx$ that results in an undesired or unpredictable performance of a model. In the case of a VAE, we construct an adversarial input $\rvx^a$ as a deformation of a reference point $\rvx^r$ to satisfy the following conditions: (\textit{i}) $\rvx^a$ should be close\footnote{Either visually or in terms of the pixel values.} to $\rvx^r$; (\textit{ii}) the variational distribution $\Enc{\rvz}{\rvx}$ for the $\rvx^r$ and the $\rvx^a$ should be different, and the same should hold for the conditional likelihood $\Dec{\rvx}{\rvz}$.
Next, we define a framework that can be used to define and compare adversarial attacks and, most importantly, evaluate the VAE robustness to them. 
\paragraph{Attacks construction} We define an adversarial point as a result of the additive perturbation of the reference point,  $\rvx^{a} = \rvx^{r} + \epsilon^{*}$, where the perturbation $\epsilon^{*}$ is a solution of an optimization problem. We distinguish between the \textit{supervised attack}, when we have access to a target point, $\rvx^{t}$ and the \textit{unsupervised attack}, when $\rvx^{t}$ is not available. In the former case, we propose to solve the following optimization problem:\footnote{$ \mathrm{SKL}[p_1 , p_2] = \frac{1}{2}\mathrm{KL}[p_1 || p_2] + \frac{1}{2}\mathrm{KL}[p_2 || p_1]$ is the symmetric version of the Kullback-Leibler divergence}
\begin{equation}\label{eq:supervised_attack}
    \rvepsilon^* = \argmin_{\|\rvepsilon\|\leq 1}  \mathrm{SKL}[\Enc{\rvz}{\rvx^{r}+\rvepsilon} , \Enc{\rvz}{\rvx^{t}}].
\end{equation}
In the unsupervised case, we formulate the following optimization problem:
\begin{align}\label{eq:unsupervised_attack}
    &\rvepsilon^* = \argmax_{\|\rvepsilon\|\leq 1}  \Delta[\Enc{\rvz}{\rvx^{r}+\epsilon} , \Enc{\rvz}{\rvx^{r}}] = \argmax_{\|\rvepsilon\|\leq 1} \| \mathbf{J}_{\rvx^{r}}^{q} \epsilon\|_{2}^{2}
\end{align}
where $\mathbf{J}_{\rvx^{r}}^{q}$ is the Jacobian of $\Enc{\rvz}{\rvx^{r}}$ at point $\rvx^{r}$. See Appx. \ref{appendix:delta} for the details for this objective.

\paragraph{Robustness measures} It is important to define proper quantitative measures that will reflect our expectations from adversarial attacks discussed in Section \ref{sect:adversarial_attacks}. First, we focus on measuring differences in the latent spaces. For this purpose, we propose to use the following measure:
\begin{equation}
    \Omega = \sum_{\rvx^{r}} \sum_{\rvx^{a}|\rvx^{r}} \mathrm{SKL}[\Enc{\rvz}{\rvx^{a}} , \Enc{\rvz}{\rvx^{r}}] ,
\end{equation}
where the value of $\epsilon$ is a solution from either (\ref{eq:supervised_attack}) or (\ref{eq:unsupervised_attack}). 

Further, we would like to measure similarities between $\rvx^{r}$ and $\rvx^{a}$, and their reconstructions. Here, we propose to use the Multi-Scale Structural Similarity Index Measure ($\mathrm{MSSSIM}$) \citep{wang2003multiscale} which is a perception-based measure calculated at different scales, $\mathrm{MSSSIM} \in [0, 1]$:
\begin{itemize}[leftmargin=*]
    \item $\mathrm{MSSSIM}[\rvx^{r}, \rvx^{a}]$: the similarity between a reference and the corresponding adversarial input;
    \item $\mathrm{MSSSIM}[\widetilde{\rvx}^{r}, \widetilde{\rvx}^{a}]$: the similarity between reconstructions of $\rvx^{r}$ and the corresponding $\rvx^{a}$;
\end{itemize}
A successful adversarial attack would have large $\mathrm{MSSSIM}[\rvx^{r}, \rvx^{a}]$ (close to 1) and small $\mathrm{MSSSIM}[\widetilde{\rvx}^{r}, \widetilde{\rvx}^{a}]$. Moreover, for \textit{supervised attacks} we will measure similarity between reconstructions of the target and the adversarial image, $\mathrm{MSSSIM}[\widetilde{\rvx}^{t}, \widetilde{\rvx}^{a}]$. Large value of the latter would indicate a successful supervised attack.

%
%

\section{Experiments} \label{sec:experiments}
In this section we consider attacking a 1-level VAE trained on the Fashion MNIST dataset \citep{xiao2017fashion} and NVAE trained on the CelebA dataset \citep{liu2015faceattributes}. All the metrics are averaged over the reference, target, and adversarial inputs. In Appx.  \ref{appendix:experiments} we provide details on the selection of the reference and target points for both datasets. 
\subsection{VAE and \texorpdfstring{$\beta$-VAE}.}
\begin{figure}[t]
\begin{center}
\vskip -10pt
\centering
\includegraphics[width=0.95\linewidth]{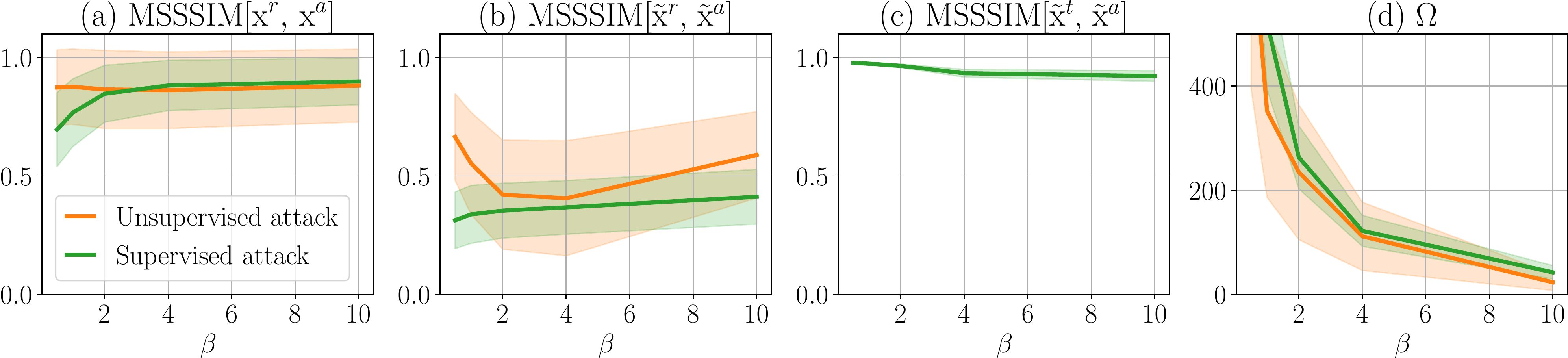}
\end{center}
\vskip -5pt
\caption{Robustness results for $\beta$-VAEs trained on Fashion MNIST dataset.}
\vskip -10pt
\label{fig:fmnist_res}
\end{figure}

We start with the experiments on the VAE with one level of latent variables. We train both VAE and $\beta$-VAE \citep{higgins2016beta}. The latter weights the KL-term in eq. (\ref{eq:elbo}) with $\beta > 0$. It is said that the larger values of $\beta$ encourage disentangling of latent representations \citep{chen2018isolating} and improve the model robustness as observed by \cite{camuto2020towards}.  In Appx. \ref{appendix:beta_vae} we provide details on the architecture, optimization, and results on the test dataset for VAE trained with different values of $\beta$. 
We observe that optimal value in terms of NLL is $\beta=1$. Larger values of $\beta$ are supposed to improve robustness in exchange for the reconstruction quality.
\begin{wrapfigure}{l}{0.46\textwidth}
\vskip 5pt
     \begin{subfigure}{0.49\linewidth}
         \centering
         \includegraphics[width=\textwidth]{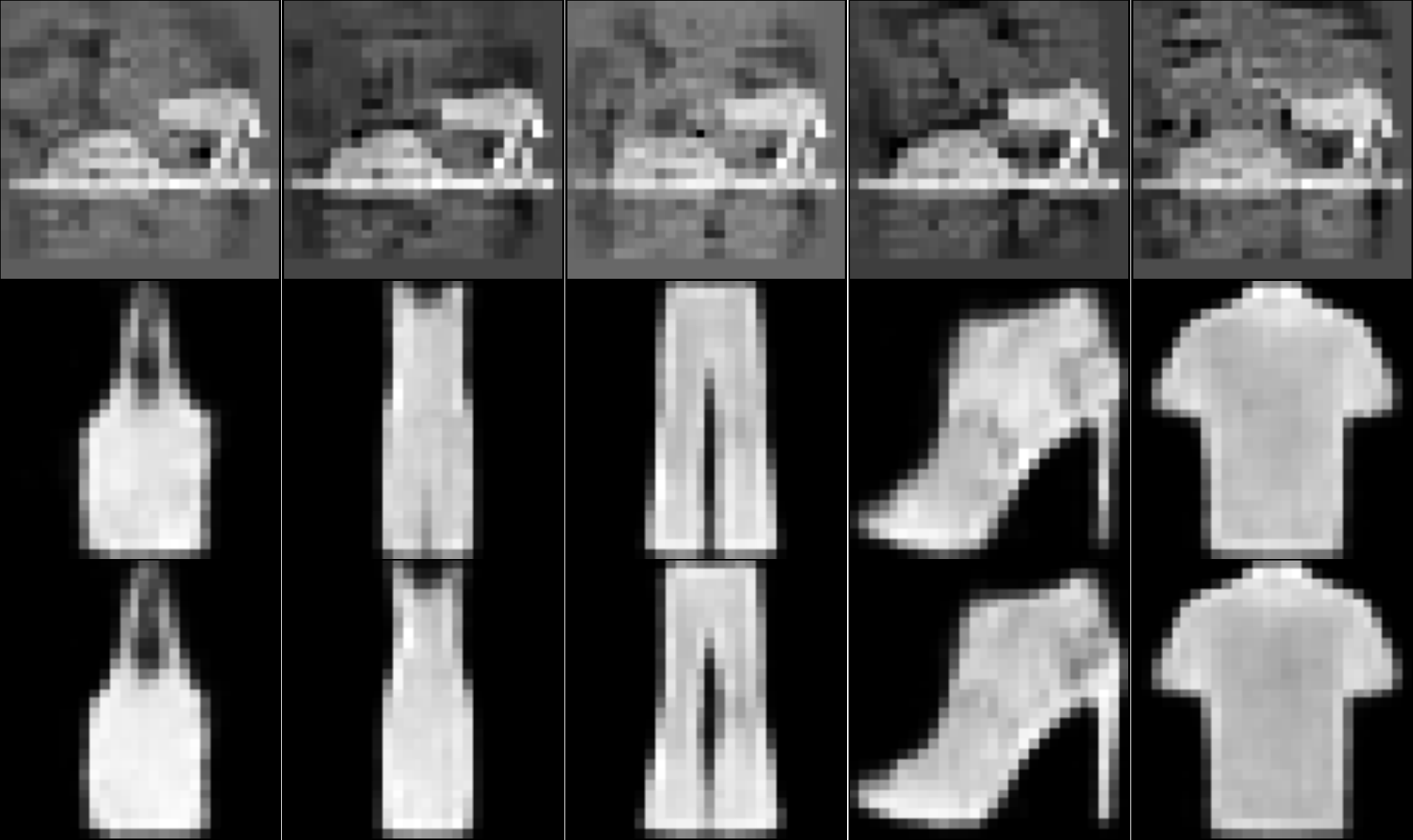}
         \caption{$\beta$ = 1}
         \label{fig:supervised_1}
     \end{subfigure}
     \begin{subfigure}{0.49\linewidth}
         \centering
         \includegraphics[width=\textwidth]{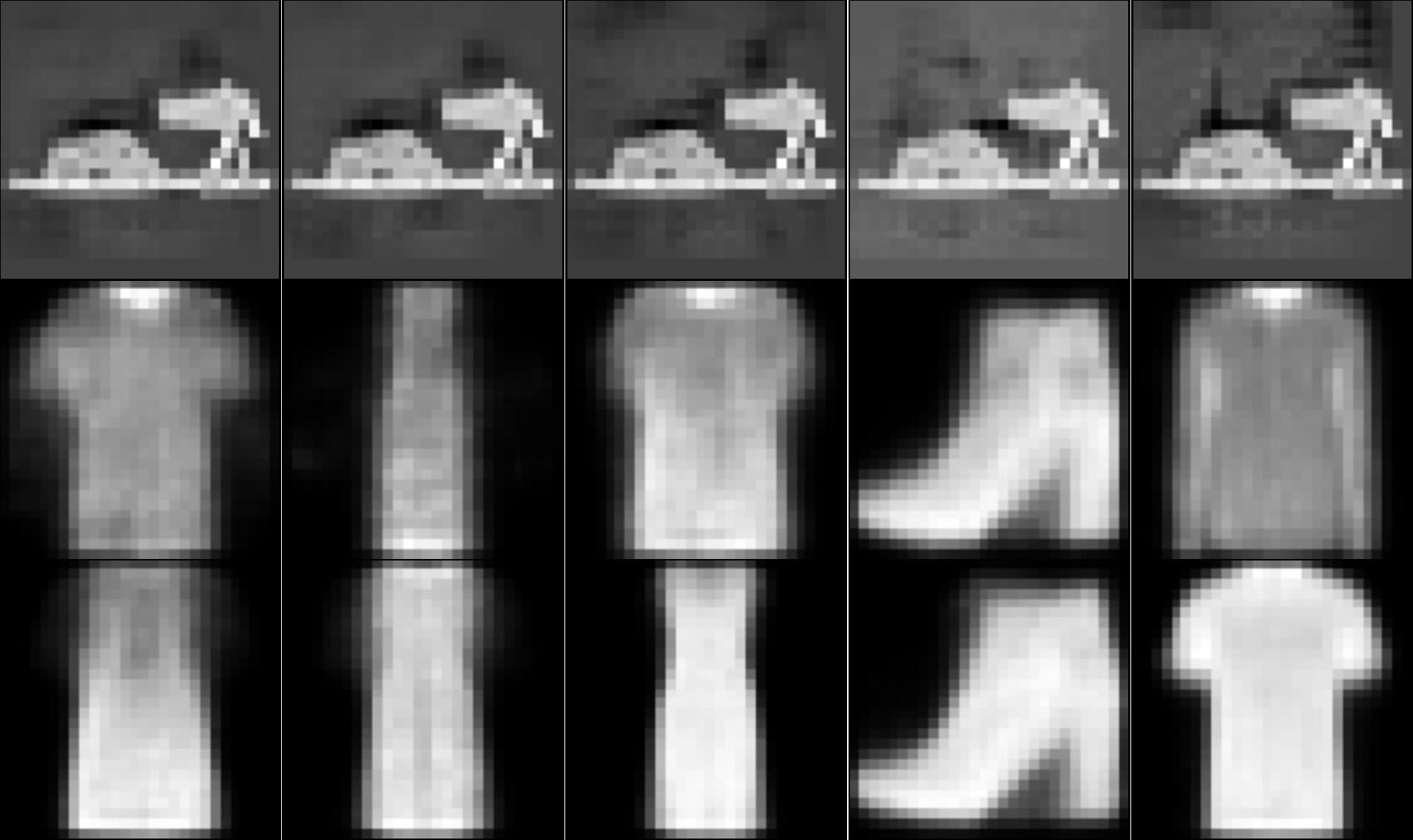}
         \caption{$\beta$ = 10}
         \label{fig:supervised_10}
     \end{subfigure}
     \caption{Supervised attack: adversarial inputs (\textit{row 1}), their reconstructions (\textit{row 2}) and reconstructions of the corresponding target points (\textit{row 3}).}
     \label{fig:sup_samples}
\vskip -10pt
\begin{center}
  \begin{subfigure}{0.38\linewidth}
         \centering
         \includegraphics[width=\textwidth]{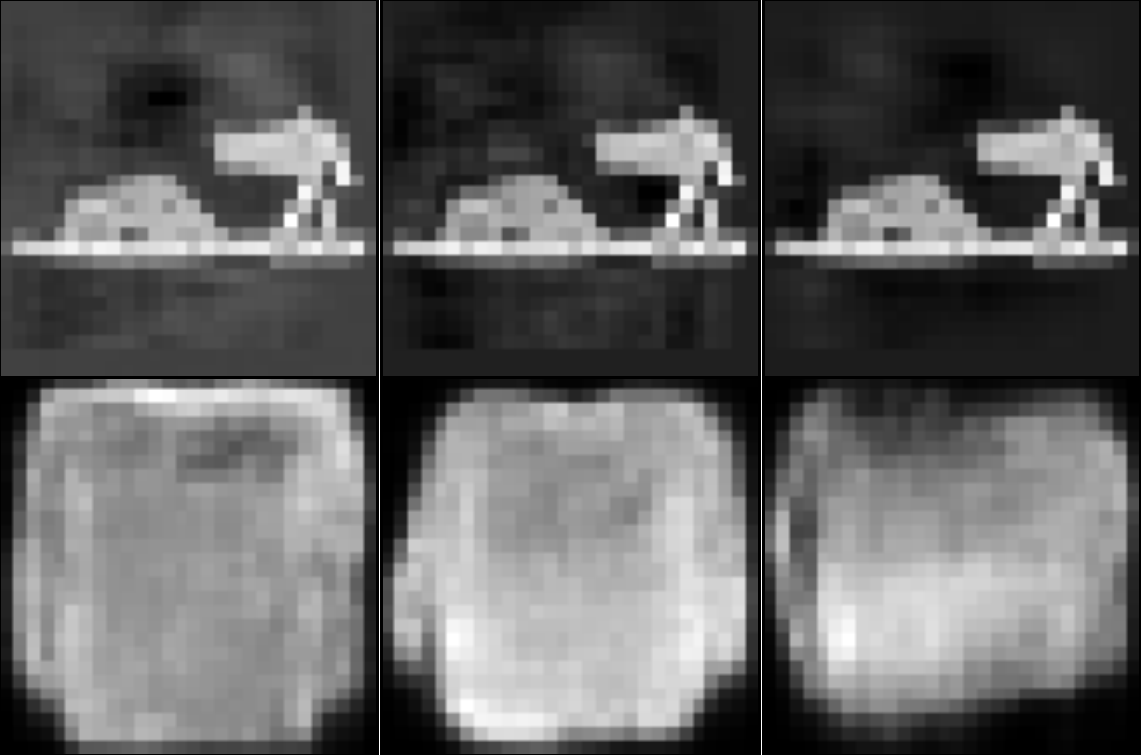}
         \caption{$\beta$ = 1}
         \label{fig:unsupervised_1}
     \end{subfigure}
     \begin{subfigure}{0.38\linewidth}
         \centering
         \includegraphics[width=\textwidth]{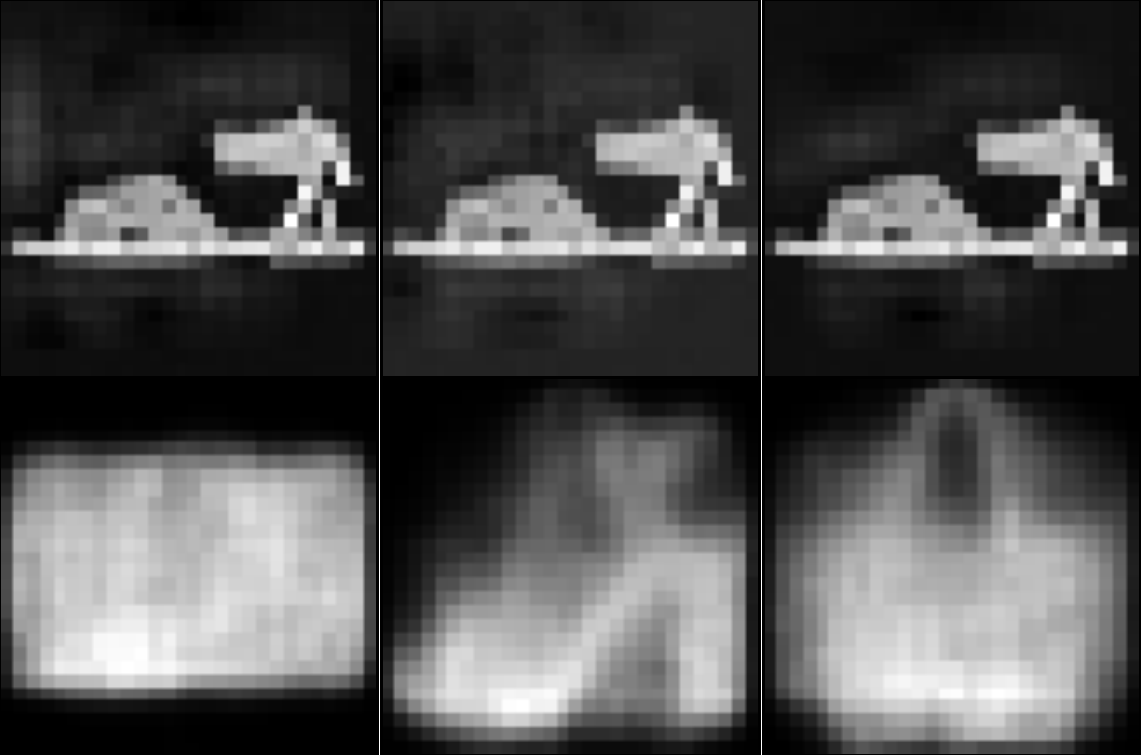}
         \caption{$\beta$ = 10}
         \label{fig:unsupervised_10}
     \end{subfigure}
      \caption{Unsupervised attack: adversarial inputs (\textit{row 1}) and their reconstructions (\textit{row 2}).}
      \label{fig:unsup_samples}
\end{center}
\vskip -35pt
\end{wrapfigure}
\paragraph{Supervised attack} We train supervised attacks using eq. (\ref{eq:supervised_attack}). Figure \ref{fig:fmnist_res} depicts result for different values of $\beta$. We observe that robustness of the encoder increases if measured by distance between adversarial and reference point in the latent space ($\Omega$). On the other hand, we still observe that in terms of reconstructions the adversarial inputs are closer to the target points than to the reference (plot (b) ad (c)).  Moreover, we do not observe higher distortion levels, that is, adversarial inputs itself are still close to the reference (plot (a)). In Figure \ref{fig:sup_samples} we provide examples of adversarial inputs for a single reference point and 5 different targets. 

\paragraph{Unsupervised attack} We present results for the unsupervised attacks trained with eq. (\ref{eq:unsupervised_attack}) in Figure \ref{fig:fmnist_res} and examples of learned adversarial inputs for a single reference point in Figure \ref{fig:unsup_samples}. We observe behavior similar to supervised tasks, where even for large values of $\beta$ we can construct a successful adversarial attack. 

\subsection{Hierarchical VAE: NVAE}
In this section, we explore the robustness of deep hierarchical VAE. It was also studied in \cite{Willetts2019-mu}, where authors notice that pure hierarchical VAE with up to 5 levels of latent variables are not gaining any robustness.
We construct a supervised adversarial attack for NVAE \citep{Vahdat2020-xe}, a recently proposed VAE with state-of-the-art performance in terms of image generation. We use a model trained on CelebA dataset using the official NVAE implementation\footnote{The code and model weight were taken from \texttt{https://github.com/NVlabs/NVAE}}.

We notice that to effectively attack a hierarchical VAE model, one has to consider only higher-order levels of latent variables, e.g. $\{z_{L-k_A}, z_{L-k_A+1}, \dots, z_L\}$. That being said, we formulate adversarial attacks using eq. (\ref{eq:supervised_attack}), with $\Enc{\rvz}{\rvx} = \prod_{i = L-k_A}^{L} \Enc{\rvz_i}{\rvx, \rvz_{>i}}$,
where $k_A$ is number of latent variables considered during attack construction. We assume that lower-order latent variables are responsible for the specific details of an image and are less useful to learn an adversarial input. A similar approach was suggested in \cite{Maaloe2019-bp} for anomaly detection. They use modified ELBO $\mathcal{L}^{>k}$, where they use prior instead of variational approximation for the first $k$ latent variables. 

\begin{figure}
\centering
\begin{minipage}{.57\textwidth}
\begin{center}
     \begin{subfigure}{0.1\linewidth}
         \centering
         \includegraphics[width=\textwidth]{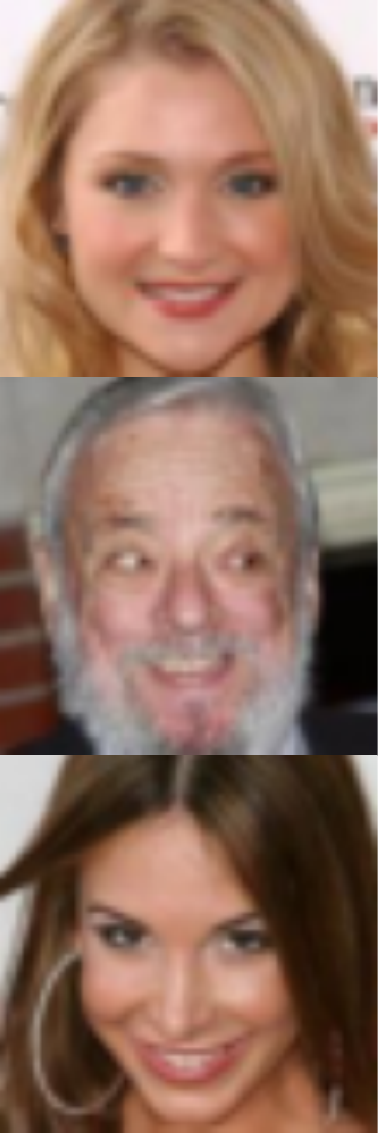}
         \caption{$\rvx^t$}
         \label{fig:nvae_trg}
     \end{subfigure}
     \begin{subfigure}{0.2\linewidth}
         \centering
         \includegraphics[width=\textwidth]{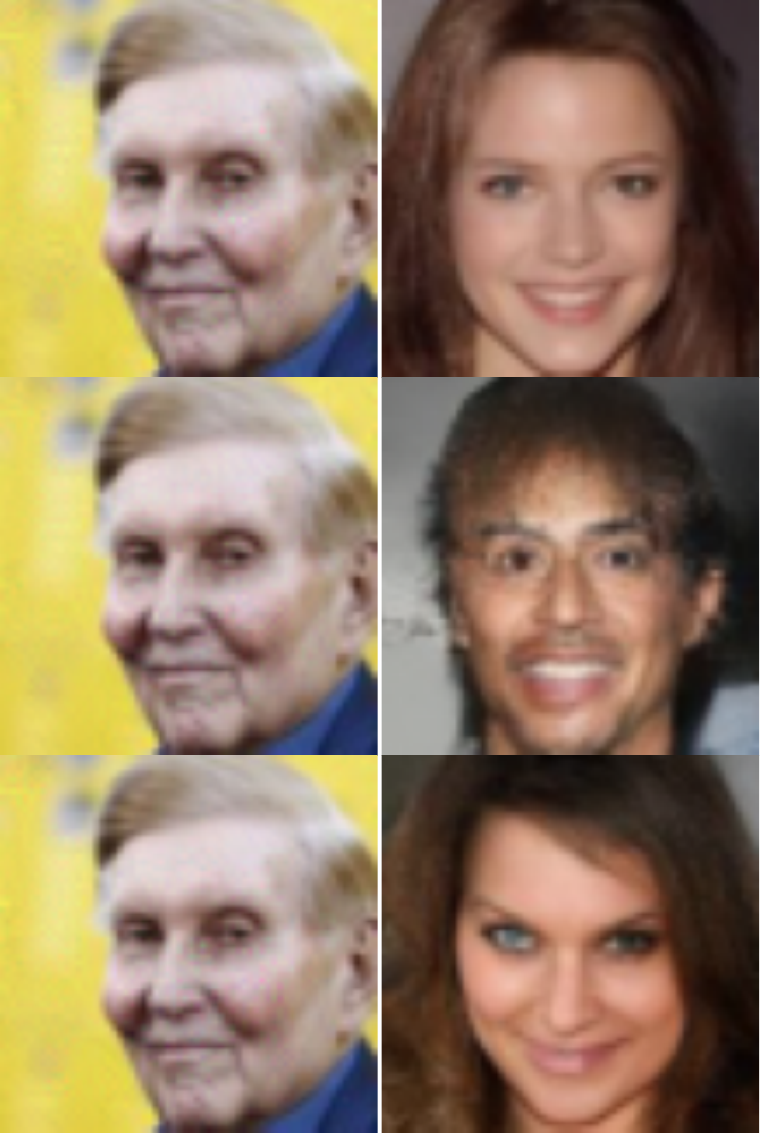}
         \caption{$k_A = 1$}
         \label{fig:nvae_1}
     \end{subfigure}
     \begin{subfigure}{0.2\linewidth}
         \centering
         \includegraphics[width=\textwidth]{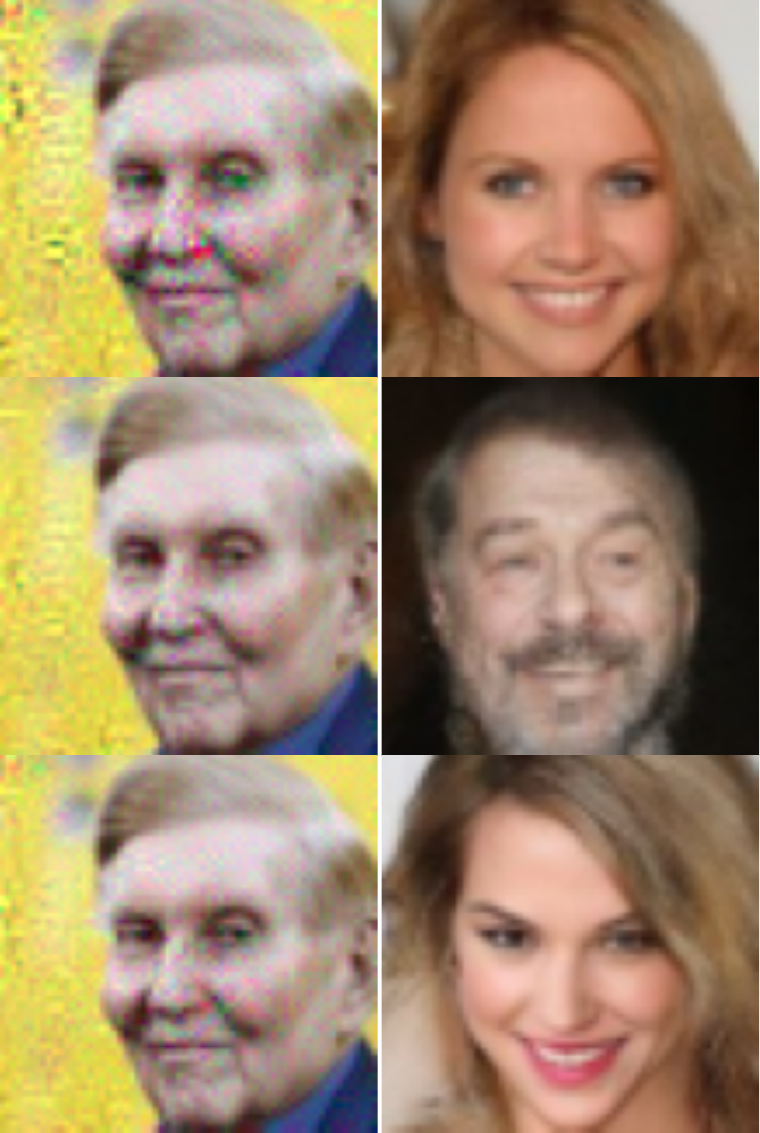}
         \caption{$k_A = 2$}
         \label{fig:nvae_2}
     \end{subfigure}
      \begin{subfigure}{0.2\linewidth}
         \centering
         \includegraphics[width=\textwidth]{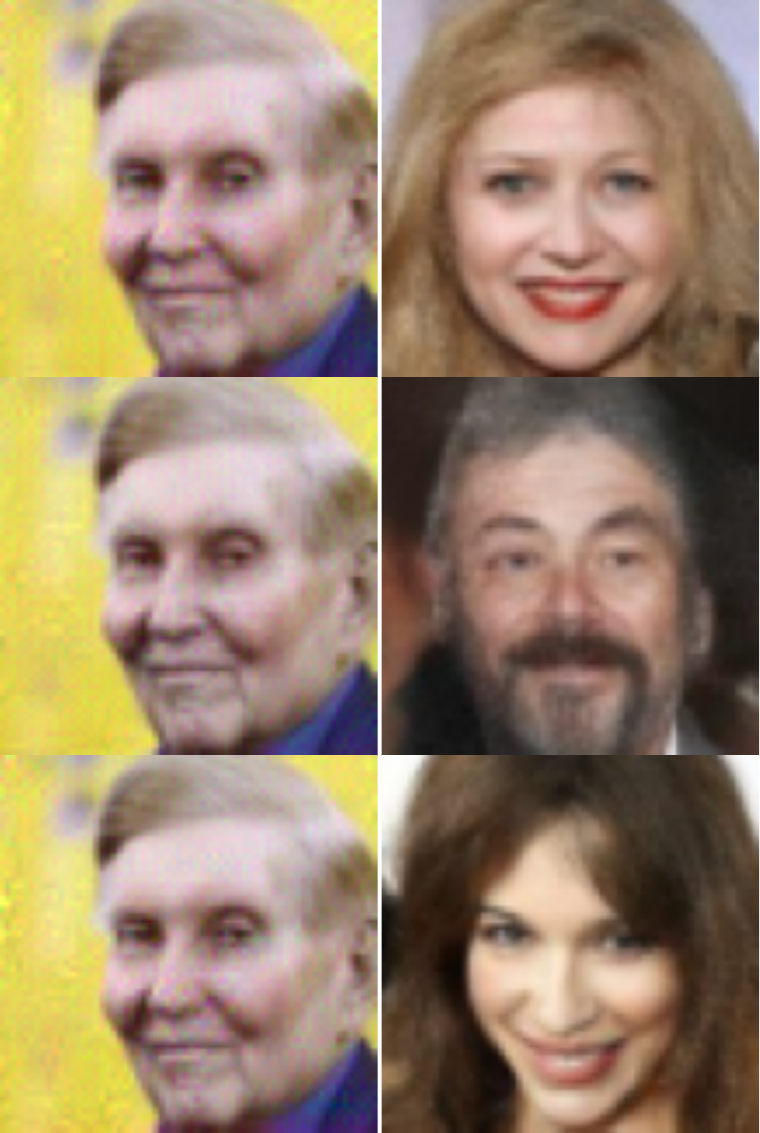}
         \caption{$k_A = 4$}
         \label{fig:nvae_4}
     \end{subfigure}
      \begin{subfigure}{0.2\linewidth}
         \centering
         \includegraphics[width=\textwidth]{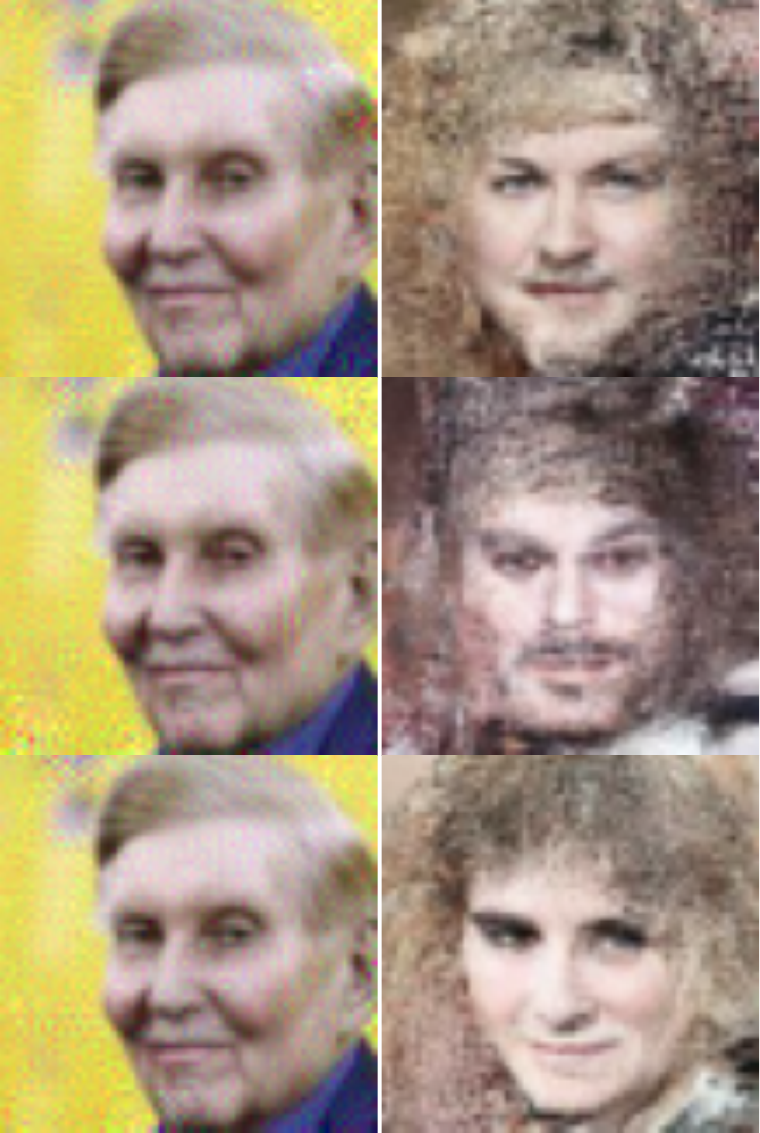}
         \caption{$k_A = 8$}
         \label{fig:nvae_8}
     \end{subfigure}
     \caption{Supervised attacks on NVAE. We plot target images in (a). In (b) - (e) we plot adversarial inputs (\textit{column 1}) and their reconstructions (\textit{column 2}). $k_A$ stands for number of top level latent variables considered while learning $\rvx^a$. We observe that adversarial reconstructions are able to mimic such high level features of the target as face orientation, hairstyle and smile.}
     \label{fig:sup_samples_nvae}
\end{center}
\end{minipage}%
\begin{minipage}{.03\textwidth}
\hskip 2pt
\end{minipage}%
\begin{minipage}{.39\textwidth}
\vskip -10pt
  \begin{center}
\includegraphics[width=1.05\linewidth]{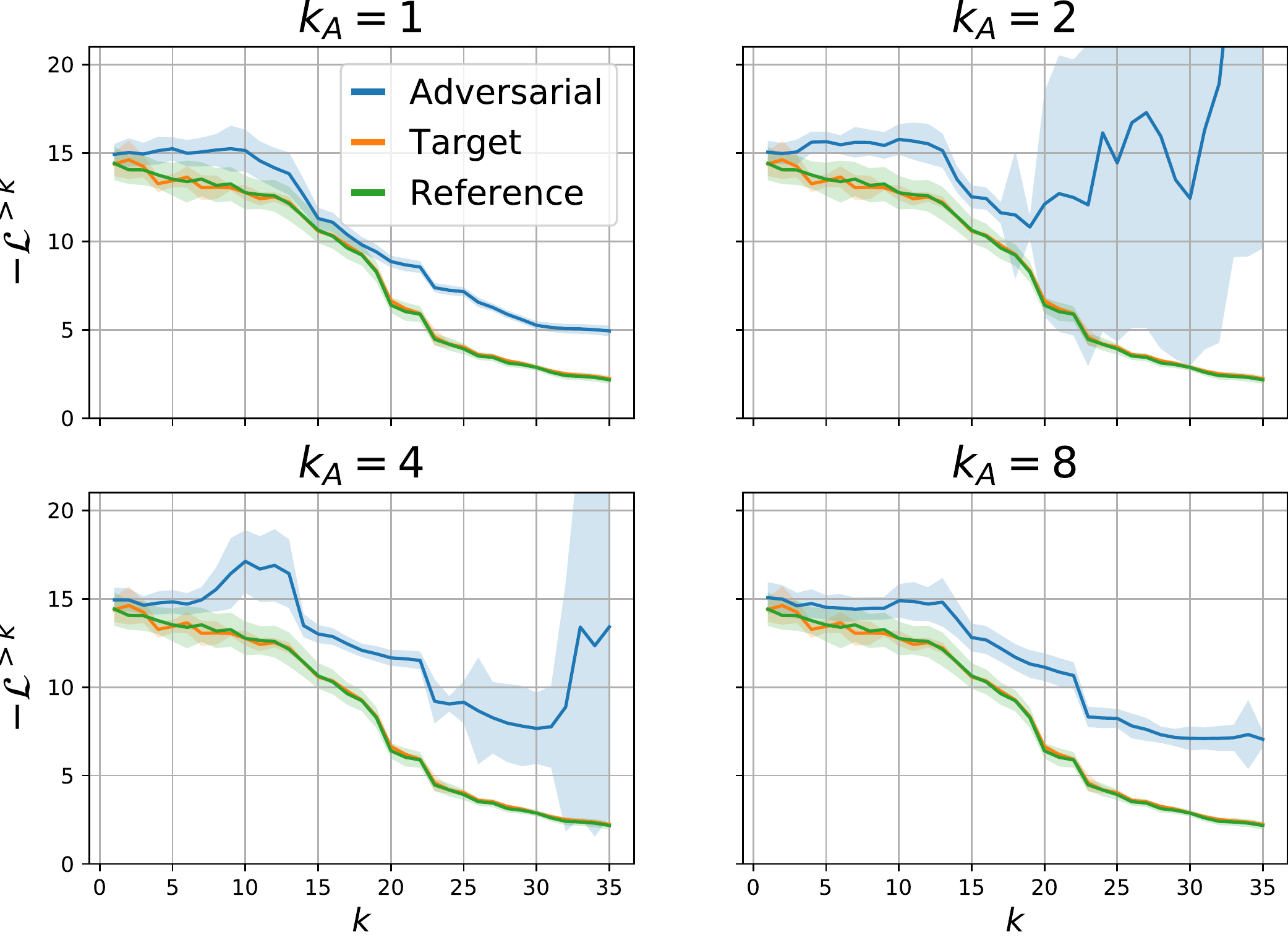}
\end{center}
\vskip -10pt
\caption{Negative ELBO $\mathcal{L}_k$ from \cite{Maaloe2019-bp}. Higher values indicate anomaly.}
\label{fig:nvae_elbo}
\end{minipage}
\end{figure}

In Table \ref{tab:nvae_res} we present numerical results for attacks with $k_A = \{1, 2, 4, 8\}$. We also plot examples of the learned adversarial inputs with their reconstructions and corresponding target images in Figure \ref{fig:sup_samples_nvae}. In Figure \ref{fig:nvae_elbo} we plot $- \mathcal{L}^{>k}$ from \cite{Maaloe2019-bp} for all possible values of $k$. We see that curves for adversarial inputs are always above those for the real images from the dataset (either target or reference ones). Moreover, according to our metrics, we are able to obtain an adversarial input that has reconstructions close to the target images rather than to the reference ones. This makes us question the robustness of the hierarchical models to adversarial attacks.

\begin{table}[t]
\vskip -0pt
\caption{Supervised adversarial attacks on NVAE in terms of proposed metrics.}
\vskip -9pt
\label{tab:nvae_res}
\begin{center}
\begin{adjustbox}{max width=.9\linewidth}{}
\begin{tabular}{l|cccc}
\toprule
\small $k_A$ &  \small $\mathrm{MSSSIM}[\rvx^{r}, \rvx^{a}]$ & \small $\mathrm{MSSSIM}[\widetilde{\rvx}^{r}, \widetilde{\rvx}^{a}]$ & \small $\mathrm{MSSSIM}[\widetilde{\rvx}^{t}, \widetilde{\rvx}^{a}]$ & \small $\Omega$ \\\midrule
1 & 0.99& 0.25& 0.51 & 270\\
2 & 0.97& 0.25& 0.65 & 281 \\
4 & 0.98& 0.30& 0.55 & 328\\
8 & 0.99& 0.46& 0.42 & 803\\ \bottomrule
\end{tabular}
\end{adjustbox}
\end{center}
\end{table}

\section{Conclusion}
In this work, we have explored the robustness of VAEs to adversarial attacks. We have suggested metrics that are easily interpretable. We have used these metrics as well as the proposed definition of supervised and unsupervised attacks to show that VAE and $\beta$-VAE are prone to adversarial attacks. Moreover, we were able to attack deep hierarchical VAE to produce high quality adversarial inputs and reconstructions.

\section*{Acknowledgements}
This research was (partially) funded by the Hybrid Intelligence Center, a 10-year programme  funded by the Dutch Ministry of Education, Culture and Science through the Netherlands  Organisation for Scientific Research, https://hybrid-intelligence-centre.nl.

\bibliographystyle{iclr2021_conference}
\bibliography{biblio}
\newpage
\appendix
\section*{Appendix}
\section{\texorpdfstring{$\Delta$} objective for unsupervised attacks}
\label{appendix:delta}

We start with defining the adversarial objective as a difference between the mean evaluated on the adversarial input ($\rvx^{a} = \rvx^{r} + \epsilon$) and the mean evaluated on the reference point, namely:
\begin{equation} \label{eq:unsup_def}
    \widetilde{\Delta}[\Enc{\rvz}{\rvx_{r}+\epsilon} , \Enc{\rvz}{\rvx_{r}}] 
    \stackrel{df}{=} \|\mu(\rvx^{r} + \epsilon ) - \mu(\rvx^{r}) \|_{2}^{2} .
\end{equation}

Since we consider a small perturbation $\epsilon$, we assume that it is reasonable to use linear approximation of the change in $\mu$. That is, we can approximate $\mu(\rvx^{r} + \epsilon )$ using on its value in $\rvx^{r}$ and Jacobian evaluated at $\rvx^{r}$.
\begin{equation}
    \mu(\rvx^{r} + \epsilon ) \approx \mu(\rvx^{r}) + \mathbf{J}_{\rvx^{r}}^{q} \epsilon^{\top}
\end{equation}

If we plug that into the \eqref{eq:unsup_def}, we get an objective:
\begin{align}
    \widetilde{\Delta}[\Enc{\rvz}{\rvx_{r}+\epsilon} , \Enc{\rvz}{\rvx_{r}}] 
    &\stackrel{df}{=} \|\mu(\rvx^{r} + \epsilon ) - \mu(\rvx^{r}) \|_{2}^{2} \\
    &\approx \|\mu(\rvx^{r}) + \mathbf{J}_{\rvx^{r}}^{q} \epsilon^{\top} - \mu(\rvx^{r}) \|_{2}^{2} \\
    &= \| \mathbf{J}_{\rvx^{r}}^{q} \epsilon^{\top} \|_{2}^{2} 
\end{align}
Note, that matrix $\mathbf{J}_{\rvx^{r}}^{q}$ does not depend on $\epsilon$. Therefore, we only need to compute it once for a given reference point.

\section{Details of the experiments}
\subsection{$\beta$-VAE} \label{appendix:beta_vae}

\paragraph{Architecture} We use fully convolutional architecture with latent dimension 128. In Table \ref{tab:fmnist_arch} we provide detailed scheme of the architecture. We use $\texttt{Conv(3x3, 1->32)}$ to denote convolution with kernel size $\texttt{3x3}$, $\texttt{1}$ input channel and  $\texttt{32}$ output channels. We denote stride of the convolution with $\texttt{s}$ and padding with $\texttt{p}$. The same notation applied for the transposed convolutions ($\texttt{ConvTranspose}$).

\begin{table}[h]
\caption{Convolutional architecture for VAE.}
\vskip -4mm
\label{tab:fmnist_arch}
\begin{center}
\begin{tabular}{@{}lll@{}}
\toprule
                   & Encoder & Decoder  \\  \midrule
& \texttt{Conv(3x3, 1->32, s=1, p=1)} & \texttt{ConvTranspose(3x3,128->256,s=1,p=0)} \\
& \texttt{ReLU()}& \texttt{ReLU()} \\
& \texttt{Conv(5x5, 32->64, s=2, p=0)} & \texttt{ConvTranspose(3x3,256->128,s=2,p=0)}\\
& \texttt{ReLU()} & \texttt{ReLU()}\\
& \texttt{Conv(5x5, 64->128, s=2, p=0)} &\texttt{ConvTranspose(4x4,128->64,s=2,p=0)} \\
& \texttt{ReLU()} &  \texttt{ReLU()} \\
& \texttt{Conv(3x3,128->256,s=2,p=1)} & \texttt{ConvTranspose(4x4,64->1,s=2,p=1)} \\
& \texttt{ReLU()} &  $\mu_x \leftarrow$  \texttt{Sigmoid()}\\
& $\mu_z \leftarrow$  \texttt{Conv(3x3,256->128,s=2,p=1)} &  \\
& $\log \sigma^2_z \leftarrow$  \texttt{Conv(3x3,256->128,s=2,p=1)}& \\ 
\bottomrule
\end{tabular}
\end{center}
\end{table}

\paragraph{Optimization} We use Adam to perform the optimization. We start from the learning rate $5e-4$ and reduce it by the factor of 0.9 if the validation loss does not decrease for 10 epochs. We train a model for 500 epochs with the batch size 256. 

\begin{table}[h]
\caption{Test performance of the $\beta$-VAE with different values of $\beta$.
Negative loglikelihood is estimated with importance sampling as suggested in \citep{burda2015importance}
} 
\vskip -4mm
\label{tab:fmnist_res}
\begin{center}
\begin{tabular}{l|ccc}
\toprule
$\beta$ & \small $-\log p(\rvx)$    & \small $\KL{\Enc{\rvz}{\rvx}}{p(\rvz)}$   \\\midrule
.5  & 234.9 & 22.5             \\
1    & \textbf{233.9} & 15.1 \\
2    & 235.5 & 10.2               \\
4    & 239.0 & 6.8                \\
10   & 250.6 & \textbf{3.9}    \\ \bottomrule
\end{tabular}
\end{center}
\end{table} 

\paragraph{Results}
In Table \ref{tab:fmnist_res} we report negative log-likelihood (NLL) of the VAE with different values of parameter $\beta$. We observe that the optimal value in terms of NLL is $\beta=1$. Larger values of $\beta$ are supposed to improve robustness in exchange for the reconstruction quality.

\subsection{Adversarial Attack} \label{appendix:experiments}
In all the experiments we randomly select reference and target points from the test dataset. For the Fashion MNIST, we also ensure that the resulting samples are properly stratified --- include an even number of points from each of the classes. 

For supervised attacks, we learn one adversarial input for each possible pair of the target and reference point. For unsupervised attack we train 6 adversarial inputs for each reference point since we have noticed, that different initialization results in different adversarial inputs. In Table \ref{tab:setup} we provide the summary of the total number of points considered. 

\begin{table}[h]
\caption{Number of reference, target and adversarial points considered for adversarial attacks.}
\vskip -4mm
\label{tab:setup}
\begin{center}
\begin{tabular}{llccc}
\toprule
 & Dataset &\# of reference points   & \# of target points & \# of adversarial points   \\\midrule
Unsupervised & Fashion MNIST &50 & --- & 300 \\
Supervised  & Fashion MNIST &50 &   10 & 500 \\
Supervised  & CelebA & 15 &   3 & 45 \\\bottomrule
\end{tabular}
\end{center}
\end{table}

\end{document}